\documentclass[a4paper,11pt]{article}

\usepackage[margin=1in]{geometry}

\usepackage{amsmath}
\usepackage{graphicx}

\title {A New Theory of the Dark Photon and Its Preliminary Experimental Verification}

\author{Weinan Wang}

\usepackage{chngcntr}
\counterwithout{equation}{section}

\begin{document}

\maketitle

\begin{abstract}
\quad \textbf{Abstract:} The dark photon defined in this paper is fundamentally different from the dark photon in the traditional kinetic mixing framework. According to the principle of energy conservation, a complete moving photon is defined in this work as a binary composite entity consisting of a dark photon and a light-wave component. The dark photon embodies the particle nature of light, while the light-wave embodies the wave nature of light. When the light source is turned off or the light-wave component is absorbed by matter, the complete photon loses its light-wave component and transforms into a pure dark photon state. In conventional high-energy physics, dark photons from kinetic mixing scenarios exhibit no direct electromagnetic coupling and cannot be detected through experiments. In distinct contrast, the dark photon proposed in this work possesses intrinsic particle properties of light, enables direct electromagnetic interaction, and can induce the photoelectric effect upon incidence on a metal surface. Based on this self-consistent definition, a systematic theory of dark photons is established, which provides reasonable physical interpretations for several currently controversial optical phenomena. The experimental scheme in this paper is designed strictly based on the above theoretical premise. Preliminary tabletop experimental results verify the rationality and feasibility of the proposed dark photon theory.
\end{abstract}

\section{Introduction}

\quad\quad The dark photon model and theory proposed in this paper are entirely new. They are developed upon extensive study of mainstream dark photon theories yet differ fundamentally from those conventional theoretical frameworks.

In mainstream dark photon theories, dark photons are conventionally defined as hypothetical U(1) gauge bosons beyond the Standard Model and are widely regarded as promising dark matter candidate particles. In recent years, dark photon research has advanced rapidly in both theoretical development and experimental detection techniques. Andrea Caputo et al. (2021) and Andrea Caputo \& Rouven Essig (2026) systematically summarized the theoretical foundations, kinetic mixing mechanisms, experimental constraints over the full mass range, and long term research outlook for conventional dark photon models, constructing a comprehensive reference framework for mainstream investigations in this field [1,2]. Berlin et al. (2023) reviewed landmark technological breakthroughs in axion and dark photon detection, analyzed the application scope and intrinsic strengths of different experimental platforms, and outlined the technical roadmap for next generation high sensitivity detectors [3]. Motivated by these mainstream theoretical frameworks, researchers have devised a variety of experimental strategies for dark photon searches, including superconducting cavity detection, accelerator collision experiments, space borne astronomical observations, and low background ground laboratory measurements [4-12]. Among these efforts, Haipeng An et al. (2025) carried out the first in orbit radio band dark photon measurement with the Parker Solar Probe, substantially tightening spatial constraints on the parameter space of low mass dark photons [13]. The theoretical origin of modern dark photon studies can be traced back to Holdom’s dual-U(1) gauge symmetry model published in 1986 [14]. This seminal work introduced kinetic mixing, the dynamical coupling mechanism governing interactions between ordinary photons and dark photons; this mechanism has thereafter served as the cornerstone for nearly all mainstream theoretical derivations and experimental search strategies.

Within the conventional paradigm, dark photons are treated as independent spin-1 gauge bosons that interact with ordinary photons solely through quantum kinetic mixing. This theoretical framework was originally motivated by the dark matter candidate puzzle. Nevertheless, after decades of theoretical development and large-scale experimental campaigns, no direct observational signature of dark photons has been uncovered within this conventional framework. Furthermore, the paradigm cannot provide an intuitive microscopic explanation for fundamental optical phenomena such as wave‑particle duality and the observer-dependent effect in double-slit interference, and no unified understanding has been achieved.

Different from conventional dark photon theories, the dark photons defined in this paper constitute a concept fundamentally distinct from those in the traditional dynamical hybrid framework. In this paper, we construct a self-contained, self-consistent theoretical framework for dark photon physics: a complete photon in motion is defined as a binary entity comprising a dark photon and a light-wave. The dark photon embodies the particle nature of photon, whereas the light-wave embodies its wave nature. Rather than being strictly confined to classical mechanics, special relativity, or standard quantum field theory, our theoretical framework establishes a closed, logically coherent system founded on the hypothesized internal composite structure of photons. In the traditional dynamical hybrid framework, dark photons do not couple directly to the electromagnetic field; hence, they cannot be detected directly. By contrast, the dark photons defined in this paper carry photon-like particle properties and can induce the photoelectric effect when incident upon a metal surface. An early prototype of this theoretical framework was first posted on arXiv in December 2024 [15]; the present manuscript further refines the theoretical derivations and provides preliminary experimental validation.

The dark photon theory proposed herein offers consistent explanations for multiple outstanding cosmological and optical anomalies that resist interpretation under the standard paradigm. It therefore opens up a low-cost and promising research avenue for exploring the nature of light, investigating dark photon physics and advancing cosmological studies.
 
\section{A new dark photon theory}

\quad\quad This paper puts forward an original, self-contained theoretical framework for dark photons. Its core physical picture originates from the composite photon model proposed by the author (see Ref.15). This model stems from a fundamental fact: physical entities cannot appear out of nowhere, nor can they vanish without a trace. When the light in a dark box with mirrored inner walls is turned off, visible light disappears immediately, and the inside of the dark box becomes pitch black instantly. This naturally raises a question: if the inner walls of the box are completely non-absorbing and non-transmitting to light, where does the light go? Therefore, this paper provides the following definition: a fully propagating photon $\gamma$ is defined as a system composed of two coupled constituents: a light-wave component $\gamma_{W}$ and a dark photon component $\gamma_{D}$. Equivalently, a free dark photon is understood as a photon that has entirely shed its light-wave component. A schematic illustration of this composite structure is presented in Figure 1. 

\begin{figure}[!htbp]
        \centering
        \includegraphics[width=0.5\linewidth]{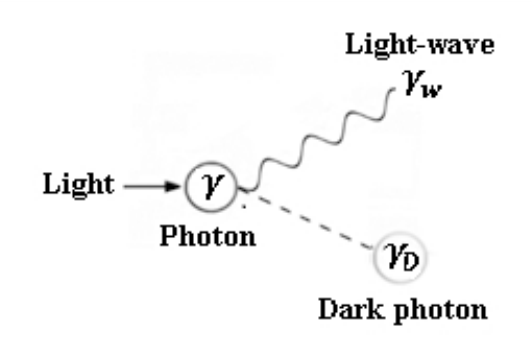}
        \caption{ Schematic diagram of the dark photon model}
        \label{fig:placeholder}
    \end{figure}

\par

Equations (1) and (2) mathematically express the internal relationship among a photon, its dark photon component, and its light-wave component:

\vspace{15pt}
\par
\begin{equation}
    \gamma\rightarrow\gamma_{D}+\gamma_{W}
\end{equation}
\begin{equation}
    \gamma_{D}\rightarrow\gamma-\gamma_{W}
\end{equation}

\par
\vspace{15pt}
\par

In equations (1) and (2), $\gamma$ denotes the fully propagating photon; $\gamma_{D}$ represents its dark photon component; and $\gamma_{W}$ stands for its bound light-wave component carried by the photon. The plus and minus signs in these equations denote component combination and separation, respectively, rather than algebraic operations.

The core idea of this model is that light is composed of numerous individual photons, and each photon consists of one dark photon and one light-wave component. When a photon loses its light-wave component, it transforms into a pure dark photon state. This definition is fundamentally different from the dark photon concept adopted in the traditional kinetic mixing model. The dark photon defined in this work retains the intrinsic particle-like properties of photons and can excite surface electrons of metals to induce the photoelectric effect. In contrast, conventional dark photons from kinetic mixing frameworks cannot couple directly to electromagnetic fields and therefore cannot produce such optical interactions.  

Based on this composite structural hypothesis, the wave-like behavior of photons directly originates from the bound light-wave component, while all particle-like characteristics stem from the dark photon component. This binary separation mechanism provides a natural and intuitive microscopic explanation for the long-debated wave-particle duality of light.

The energy-momentum relations formulated by de Broglie serve as mathematical tools for describing photon behavior under the traditional single component photon hypothesis, as expressed in Equations (3) and (4). These formulas describe the outcomes measured in conventional optical experiments and remain useful mathematical tools within the theoretical framework of this paper.

\begin{equation}
E = h\nu = pc
\label{eq:3}
\end{equation}
\begin{equation}
p = h/\lambda
\label{eq:4}
\end{equation}

In the above formulas, $E$ represents the  energy of a single photon measured in conventional optical experiments; $p$ denotes the photon’s momentum; $\nu$ stands for its frequency; $\lambda$ represents its wavelength; $h$ is the Planck constant; and $c$ is the speed of light in vacuum. 

Within the independent theoretical framework established in this paper, a complete photon consists of two components, each carrying a distinct physical attribute. This binary internal structure is a core feature of the dark photon theory. Among them, the light-wave component possesses neither rest mass nor dynamic mass, yet it independently carries wave energy, as described in Equation (5).

\begin{equation}
E_{w} = h\nu
\label{eq:5}
\end{equation}

$E_{W}$ represents exclusively the wave energy carried by the light-wave.

Complete photons propagate at the speed of light and thus possess momentum. In the dark photon model, since the light-wave have neither rest mass nor kinetic mass, the total kinetic energy of the photon system originates entirely from the dark photon component. Therefore, in this paper, the total energy of a complete photon is defined as the sum of the wave energy of the light-wave component and the particle kinetic energy of the dark photon component, as shown in Equation (6).

\begin{equation}
E_{\mathrm{total}} = E_{w} + E_{D} = h\nu + pc
\label{eq:6}
\end{equation}

In Equation (6), $E_{total}$  represents the total energy of a photon. $E_{w}$ represents the wave energy carried by the light-wave component, which can be quantified for different frequency bands using Equation (5). $E_{D}$ denotes the kinetic energy carried by the dark photon component. Within the model, the particle-like behavior of photons arises entirely from their dark photon component; thus, the kinetic energy associated with the dark photon is equivalent to that of the photon. $E_{D}$ cannot be measured directly by conventional methods and can only be determined from the momentum of photons, which can be obtained through light pressure or the Compton effect. Therefore, standard optical techniques cannot directly yield the total photon energy $E_{total}$; the total energy can only be obtained by separately measuring  $E_{w}$ and $E_{D}$ and then summing the results. Equations (1), (2), and (6) together constitute the core theoretical framework of the dark photon theory presented in this paper. 

This theoretical framework for composite dark photons strictly adheres to the principles of energy and momentum conservation.

In terms of energy conservation, the total energy of a propagating photon consists of the kinetic energy associated with its dark photon component and the wave energy carried by its light-wave component. When a photon sheds its bound light-wave component, the wave energy previously associated with that component is released as heat or radiation. Meanwhile, the kinetic energy originally attributed to the full photon is fully retained by the residual dark photon component. Thus, the total energy of the system remains conserved throughout this component separation process.

In terms of momentum conservation, according to the core tenets of dark photon theory, the light-wave component possesses neither rest mass nor dynamic mass; the entire momentum of the composite photon is therefore carried solely by the dark photon component. Consequently, the photon momentum is completely independent of the light-wave component, whether the light-wave component is bound or separated. Thus, the total momentum of the photon system remains conserved during structural decomposition.

Traditional theoretical research describes weak quantum mixing between Standard Model photons and dark photons through kinetic mixing, which gives rise to quantum oscillation phenomena. The standard Lagrangian describing this conventional interaction framework is given in Equation (7), quoted here solely for comparative discussion with the composite photon model proposed in this paper.

\begin{equation}
\mathcal{L}=-\frac{1}{4}F_{\mu\nu}F^{\mu\nu}
-\frac{1}{4}F'_{\mu\nu}F'^{\mu\nu}
-\frac{\epsilon}{2}F_{\mu\nu}F'^{\mu\nu}
+\frac{1}{2}m_{A'}^2 A'_{\mu}A'^{\mu}
\label{eq:7}
\end{equation}

In Equation (7), $\mathcal{L}$ denotes the total Lagrangian density; $F_{\mu\nu}$ and $F'_{\mu\nu}$ represent the covariant field strength tensors of ordinary photons and dark photons, respectively; $F^{\mu\nu}$ and $F'^{\mu\nu}$ are the corresponding contravariant field strength tensors; $A'_{\mu}$ and $A'^{\mu}$ are the covariant and contravariant four‑vector potentials of dark photons; $\epsilon$ stands for the kinetic mixing coupling parameter; and $m_{A'}$ is the intrinsic mass of the dark photon defined in the conventional gauge theory.

Equation (7) constitutes the core mathematical framework for mainstream direct dark photon detection experiments. The first and second terms on the right hand side represent the free kinetic terms of ordinary photons and dark photons, respectively; the fourth term is the Proca mass term for dark photons; the third term describes the standard kinetic mixing interaction, isolated as:

\begin{equation}
\mathcal{L}_{\mathrm{mix}}=\frac{\epsilon}{2}F_{\mu\nu}F'^{\mu\nu}
\label{eq:8}
\end{equation}

The coupling parameter $\epsilon$ quantifies the interaction strength between independent gauge bosons in conventional theory, and its extremely small value indicates an ultra-weak coupling between ordinary photons and dark photons, as well as a low probability of mutual conversion. This mixing mechanism describes interactions between two fully independent particle species, distinct from the internal binding interaction between dark photons and light-wave within a single composite photon in the model proposed in this paper. Although the physical nature of the coupling is fundamentally different, the low conversion probability implied by the tiny coupling constant in the conventional mixing framework may still provide an indirect reference. On this qualitative basis, we conjecture that, after separation of the light-wave component, an isolated free dark photon has an extremely low probability of reassembling into an intact composite photon.

Three distinct physical scenarios can trigger the separation of bound light-wave components from complete photon cores, thereby converting complete photons into free dark photons: 

\textbf{(1)} Termination or disappearance of the light source: In a light field containing only a single light source, when the source is turned off or disappears, all photons emitted by it lose their bound light-wave components and become free dark photons. 

\textbf{(2)} Absorption of bound light-wave components by surrounding matter: The separated light-wave components dissipate as heat or secondary radiation, leaving behind free dark photons.  

\textbf{(3)} Collision events between photons and other particles or macroscopic objects: Impact interactions may strip the bound light-wave component away from the complete photon core, generating free dark photons.

Within the present theoretical framework, a considerable fraction of stellar-origin photons may collide with celestial bodies, cosmic dust, nebulae or interstellar gas, or be absorbed by these materials, and thereby continuously transform into free dark photons. If dark photons constitute a form of dark matter in this model, the abundant conversion of stellar-produced photons into free dark photons would lead to a steadily increasing cosmic population of dark photons, resulting in a time-dependent growth in the total dark matter density. Such an accumulation effect could qualitatively provide a possible mechanism underlying the observed accelerated cosmic expansion.

Cosmic dark photons originate predominantly from stellar radiation, whereas those produced by human industrial activities are negligible by comparison.

\section{Detection method and apparatus for dark photons}

\quad\quad In traditional dark photon research, most studies have been limited to theoretical derivations, with relatively few experimental investigations and even fewer tabletop experiments conducted indoors. Guided entirely by the composite dark photon model proposed in this paper, we develop an original detection method for dark photons and design a miniature detection apparatus; its structural schematic is shown in  Figure 2.

In Figure 2, the internal walls of the dark photon generator cuboid are fully lined with reflective mirror material to minimize light absorption inside the cavity. The light source is used to emit conventional photons, and an embedded thermometer continuously monitors temperature variations within the cavity. The small emission aperture enables directional outflow of free dark photons after component separation; the external photoresistor serves as the signal-sensing element, and a digital resistance meter (R-meter) measures the photoresistor’s real-time resistance values. The dark photon generator is fabricated in cuboid form with physical dimensions of 220 × 160 × 100 mm. A standard digital multimeter functions as the R-meter, and all real-time experimental data are recorded via video for post-processing statistical analysis.

\begin{figure}[!htbp]
            \centering
  \includegraphics[width=0.75\linewidth]{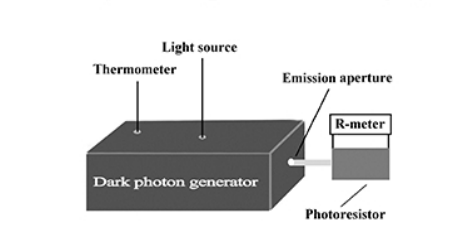}
 \caption{Schematic diagram of the self-designed dark photon detection apparatus}
  \label{fig:placeholder}
 \end{figure}
                                
The beam of light emanating from the aperture in Figure 2 schematically illustrates the direct illumination of the photoresistor during the experiment.

The overall structural design of this apparatus is based exclusively on the transformation rules of real photons and dark photons derived from the theoretical framework presented in this paper; its core measurement principle is entirely novel, and no conventional dark photon detection method serves as a reference. The complete measurement principle is elaborated below:

Owing to the near-perfect reflectivity of the mirrored inner walls, nearly all photons emitted by the light source remain trapped inside the sealed cavity without external leakage. After sustained illumination fills the cavity with photons, the light source is completely turned off. According to the transformation rules of this model, all trapped photons inside the cavity shed their bound light-wave components and convert into free dark photons, since physical entities cannot appear out of nowhere, nor can they vanish without a trace. This process establishes a significant dark photon density gradient between the high-density dark photons population inside the generator cavity and the uniform, low-density universal dark photons background field outside the device. When the small emission aperture is opened, the internal high-density dark photons flow outward toward the external photoresistor, generating measurable sensing signals and enabling detection of free dark photons.

After the light source is turned off, the number of dark photons generated inside the generator equals the number of photons previously emitted by the light source, as shown in Equation (9).

\begin{equation}
N_\gamma = N_D = n_\gamma \, t
\label{eq:9}
\end{equation}

In Equation (9), $N_\gamma$ represents the total number of photons generated by the light source inside the generator. Since the dark photon density inside the generator is the same as that outside before the aperture is sealed, $N_{D}$ denotes the increased number of dark photons produced inside the generator after the light source is turned off ,which equals the total number of photons originally emitted by the light source.$n_\gamma$ is the photon emission rate of the light source (in photons per second), and t is the illumination duration (in seconds).

Due to the increased number of dark photons inside the generator, their density is higher than the initial density of dark photons inside the generator, that is, higher than the density of dark photons in the outside environment. This creates a density gradient of dark photons between the interior and exterior of the generator. When the emission aperture on the generator opens, the high density dark photons inside escape into the external space and impinge upon the photoresistor, generating a measurable electrical signal.

The working mechanism of photoresistors relies on the internal photoelectric effect in semiconductor materials, where the core excitation process is triggered by incident photons exhibiting particle-like behavior. Within this paper’s theoretical framework, the particle-like nature of light arises exclusively from dark photon cores. When dark photons strike the sensitive surface of the photoresistor, they excite electrons within the semiconductor, generating abundant electron-hole pairs, thereby increasing the material’s conductivity and decreasing its resistance. Consequently, a sharp, measurable drop in the photoresistor’s resistance, as recorded by the R-meter, serves as an indicative signal that free dark photons have collided with the sensing surface.

A uniform, stable, low-density background dark photon field permeates the entire universe and cannot produce distinguishable detection signals under ordinary experimental conditions. The apparatus’s unique cavity design artificially creates a localized, high-density dark photon region inside the mirror-lined generator, establishing a measurable density gradient that enables effective capture of signals from free dark photons.

\vspace{10pt}

\textbf{Operating procedures }

\vspace{10pt}

\textbf{(1)} Assemble the experimental apparatus according to Figure 2. 

\textbf{(2)} Measure the initial resistance of the photoresistor without turning on the internal light source of the dark photon generator. 

\textbf{(3)} Install an aperture-stopper to seal the emission aperture of the dark photon generator, ensuring a tight seal.

\textbf{(4)} Measure the resistance of the photoresistor with the light source off and the emission aperture sealed, verifying that the deviation from the initial value is within acceptable limits.

\textbf{(5)} Turn on the light source inside the dark photon generator to fill its interior with photons; the duration of light exposure depends on the experimental design.

\textbf{(6)} Turn off the light source inside the dark photon generator. 

\textbf{(7)} After turning off the light source, remove the aperture stopper from the emission aperture. 

\textbf{(8)} After removing the aperture stopper from the emission aperture, record the resistance value of the photoresistor.

\textbf{(9)} Clean up the experimental apparatus and conclude the experiment.

\vspace{10pt}

\textbf{Main technical specifications}

\vspace{10pt}

The external dimensions of the dark photon generator: 220 × 160 × 100 mm

Emission aperture diameter: 3mm

Photoresistor resistance range: 100-200 $\mathrm{k\Omega}$  (bright value); 10$\mathrm{M\Omega}$ (dark value)

LED source: 9.5W

Thermometer resolution: 0.1℃

R-meter accuracy: $0.01\%-0.05\%$

Repeatability error (random error):$0.38\%$

\vspace{10pt}

The reported repeatability error of $0.38\%$ is obtained from calibration measurements in which the internal light source remains continuously illuminated while the emission aperture is kept open. Under this calibration condition, conventional photons directly escape from the aperture onto the photoresistor, and this dataset serves only to characterize the inherent repeatability performance of the whole measurement hardware system.

\vspace{10pt}

\textbf{Analysis of factors influencing the experiment} 

\vspace{10pt}

Various factors can influence experimental results. These factors are categorized and analyzed below, and targeted equipment and operational optimization proposals are outlined for follow-up research.

\textbf{(1)} Environmental background light interference

Photoresistor semiconductor materials are highly sensitive to subtle fluctuations in ambient light, which can independently trigger resistance changes of tens of ohms. Future follow-up experiments should be conducted in a fully light-shielded laboratory under constant, low-background illumination to minimize environmental noise.

\textbf{(2)} Influence of the sealing of the dark photon generator

The current prototype generator is assembled by bonding separate glass mirror panels, creating tiny gaps that compromise cavity sealing. During illumination, photons leak outward through these gaps; after the light source is turned off, accumulated free dark photons continuously dissipate, thereby weakening the measurable outflow signal upon aperture opening. Subsequent iterations will employ seamlessly integrated mirror cavity structures to eliminate leakage pathways.

\textbf{(3)} Uncontrollable noise introduced by manual operation

All aperture opening and data recording operations currently rely on manual execution due to preliminary experimental hardware limitations. Even after repeated operator training to standardize movement consistency, manual operation cannot achieve fully uniform timing and positioning. Moreover, the proximity of the human body can disturb the light field near the photoresistor, causing random fluctuations in the measured resistance value. The optimized experimental setup will adopt an automated electronic aperture control system and remote data acquisition to eliminate human operational interference; operators will maintain a fixed, long-distance separation from the sensing module throughout all measurement cycles.

\textbf{(4)} The influence of the distance between the generator and the photoresistor

When operated manually, a gap of at least 5 millimeters must be maintained between the dark photon generator’s emission aperture and the photoresistor’s sensitive surface, while the diameter of photoresistor’s sensitive surface is only 5 millimeters. As a result, some escaped dark photons cannot enter the sensing area at all, thereby reducing the signal amplitude. An automated electric shutter mechanism will eliminate manual operation constraints, enabling precise positioning of the photoresistor’s sensitive surface in direct proximity to the emission aperture and thereby minimizing particle loss.

\textbf{(5)} Temperature Variation Interference  

Temperature changes alter the intrinsic resistance characteristics of semiconductor photoresistor materials. Across all completed trials, the maximum internal cavity temperature rise does not exceed 3°C. Because the temperature change inside the generator is relatively small, temperature has little effect on the externally isolated photoresistor element compared to the stronger electrical signal generated by the dark photon flux overflowing from the generator onto the photoresistor. Therefore, temperature effects were not considered in this study.

\section{ Experimental results and analysis }

\textbf{(1) Visible light experimental results and analysis}

\vspace{10pt}

The visible-light source used in the experiment is a 9.5 W white LED lamp, and the sensing photoresistors are fabricated from metal sulfide semiconductor materials.

Figure 3(a) shows the averaged measurement results from five repeated parallel experiments performed on Photoresistor 1. In all the plots of this paper, $R_p$ denotes the real-time resistance of the photoresistor, t represents the time axis, $I_t$ indicates the internal illumination duration of the dark photon generator prior to light source shutdown, and $\Delta$T denotes the total temperature rise inside the generator cavity during illumination.

In the experiment, after the light source inside the dark photon emitter reaches the designed illumination duration, it is turned off and the emission aperture is opened. Under such experimental conditions, even with identical illumination parameters, the number of dark photons reaching the photoresistor in each measurement varies randomly after the light source is turned off, due to multiple random reflections of dark photons within the cavity and the continuous decay of the internal dark photon density. Consequently, two independent experiments cannot yield fully reproducible results.Therefore, Figure 3(a) presents results averaged over multiple measurements.

\begin{figure}[!htbp]
\centering
\includegraphics[width=0.49\linewidth]{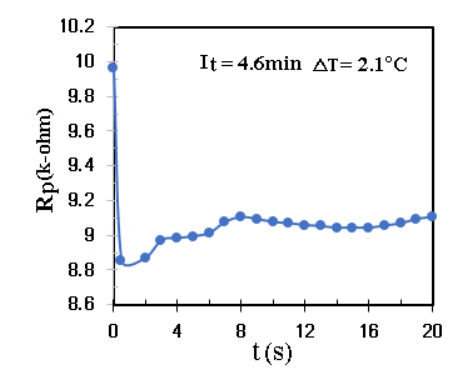}
\hspace{4mm}
\includegraphics[width=0.427\linewidth]{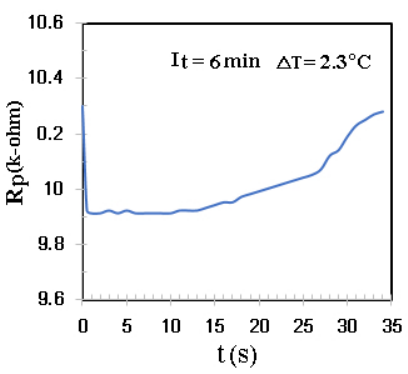}
\vspace{3pt}
\begin{minipage}{0.49\textwidth}
\centering 
\small\hspace{6mm} (a) Averaged experimental results of five parallel trials for Photoresistor 1
\end{minipage}
\hfill
\begin{minipage}{0.427\textwidth}
\centering
\small\hspace{9mm} (b) Experimental characteristic curve of photoresistor 1 
\end{minipage}
\vspace{4pt}
\caption{Experimental curves obtained from Photoresistor 1}
\label{fig:3}
\end{figure}

Figure 3(a) shows that, immediately after the internal light source is turned off and the generator’s emission aperture is opened, the resistance of photoresistor 1 decreases by an average of 736 $\mathrm{\Omega}$ over five repeated trials.The sharp drop in resistance may indicate that the photosensitive surface of the photoresistor is receiving a dark photons signal from the generator cavity. Following the initial drop in resistance, the measured resistance gradually recovered, consistent with the predicted signal decay trend as the internal dark photon density in the generator gradually equilibrates with the universal background field. This recovery behavior matches the relaxation response of photoresistors following sustained particle radiation excitation.

During the signal acquisition process, although the dark photon density in the generator continuously decreases over time, the external photoresistor still receives a gradually weakening dark photon signal. As a result, it takes a relatively long time for the photoresistor’s resistance to return to its initial value, sometimes even several tens of minutes. Figure 3(a) does not show the entire process; it only presents measurement results obtained within the first few dozen seconds. However, in some experimental runs, the photoresistor’s resistance recovery time is relatively short, as shown in Figure 3(b). This may be due to the fact that when the emission aperture of the dark photon generator opens, the intensity of dark photons emerging from the aperture is relatively high, which aligns with the photoresistor's recovery characteristics under strong light illumination.

The phenomenon shown in Figure 4(a) occurs multiple times during the experiment. A plausible explanation is that, at the moment the aperture of the dark photon generator opens, a stronger beam of dark photons escapes through it and strikes the sensitive surface of the photoresistor. This causes the photoresistor’s resistance to drop sharply and then recover to its initial value within seconds, generating a signal that resembles a negative pulse. This is consistent with the photoresistor’s recovery characteristics under strong light illumination. This negative-pulse-like signal has been repeatedly observed in the experiment and is not an interference signal.

\begin{figure}[!htbp]
\centering
\includegraphics[width=0.44\linewidth]{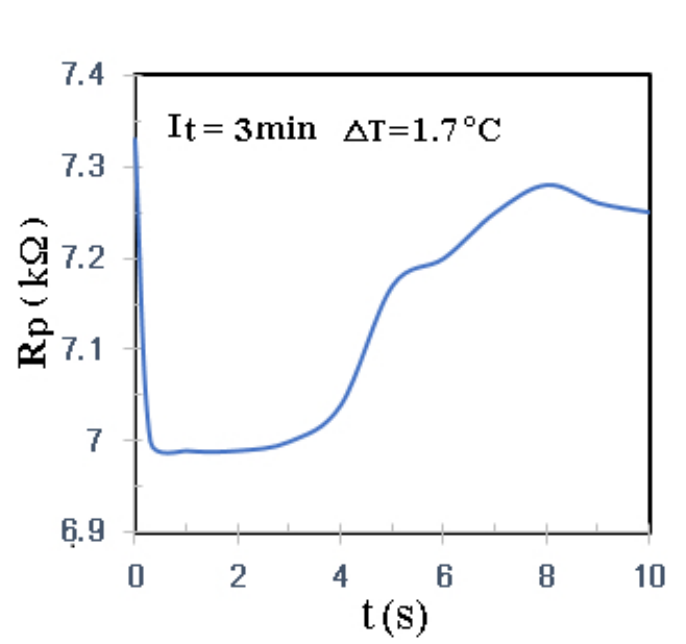}
\hspace{2mm}
\includegraphics[width=0.45\linewidth]{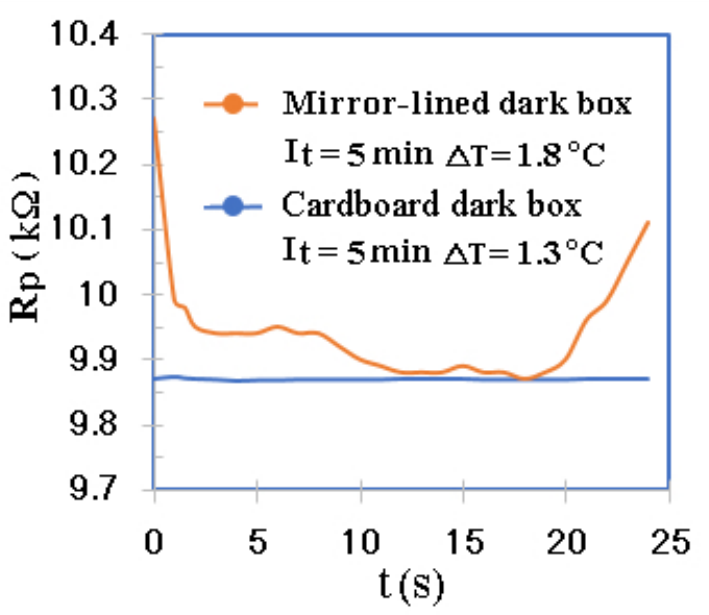}
\vspace{3pt}
\begin{minipage}{0.49\textwidth}
\centering 
\small\hspace{6mm} (a)  Experimental curve of photoresistor 2
\end{minipage}
\hfill
\begin{minipage}{0.427\textwidth}
\centering
\small\hspace{-9mm} (b)  Comparative curves of Photoresistor 11 for mirror-lined and cardboard-lined cavities
\end{minipage}
\vspace{2pt}
\caption{Experimental results for photoresistor 2 and photoresistor 11}
\label{fig:4}
\end{figure}

Figure 4(b) shows a set of comparative experiments designed to systematically verify that the mirror-lined generator cavity can effectively accumulate free dark photons after the light source is shut down. To this end, a parallel set of comparison experiments was conducted using two identically sized dark boxes with different inner-wall materials: one lined with reflective glass mirrors, the other lined with cardboard. The experimental conditions of the two groups were strictly matched. According to the predictions of the model, the mirror-lined cavity retains nearly all internal photons and subsequent free dark photons before the generator’s emission aperture opening; while cardboard-lined inner walls absorb most photons.

Figure 4(b) presents the comparative measurement results for Photoresistor 11 under both cavity configurations. After the aperture is opened, the measured curve from the mirror-lined cavity group shows a clear and distinct relaxation signal, consistent with the expected response of accumulated free dark photons exiting the cavity. In contrast, the cardboard-lined dark box group displays an approximately linear resistance curve without relaxation features, indicating no measurable outflow of free dark photons upon aperture opening. This comparative result aligns perfectly with the core predictive logic of the composite dark photon model proposed in this paper.

To further cross-validate the core predictions of the composite photon model, supplementary experiments were conducted using phototransistors as alternative high-sensitivity sensing elements. Phototransistors operate based on the same internal photoelectric effect as photoresistors but exhibit higher intrinsic signal sensitivity. Due to the high sensitivity of phototransistors, environmental background interference cannot be fully eliminated; therefore, multiple repeated measurements were performed to ensure statistically reliable conclusions.

Figure 5(a) shows experimental results for two commercial phototransistor models (3DU33 and 3DU5C), which exhibit highly consistent response characteristics. After the internal cavity light source is turned off and the emission aperture is opened, the emitter collector voltage of both phototransistors drops sharply before gradually recovering to its original value. This voltage transient matches the predicted excitation effect caused by outflowing free dark photons striking the phototransistor’s light-sensitive surface.

\begin{figure}[!htbp]
\centering
\includegraphics[width=0.50\linewidth]{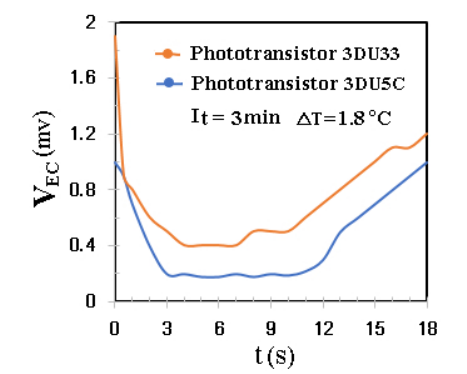}
\hspace{-1mm}
\includegraphics[width=0.49\linewidth]{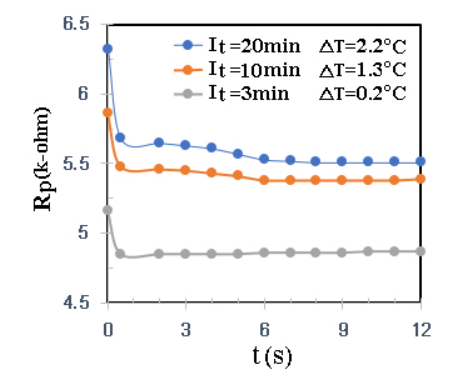}
\vspace{3pt}
\begin{minipage}{0.49\textwidth}
\centering 
\small\hspace{6mm} (a) Visible band experimental curves for two models of phototransistors
\end{minipage}
\hfill
\begin{minipage}{0.427\textwidth}
\centering
\small\hspace{-9mm} (b) Ultraviolet band experimental curves of photoresistor 3
\end{minipage}
\vspace{2pt}
\caption{Visible band results experimental for two phototransistor models and ultraviolet band
results for Photoresistor 3}
\label{fig:5}
\end{figure}

\vspace*{5pt}

\textbf{(2) Ultraviolet light experimental results and analysis}

\vspace*{5pt}

To verify whether the composite photon model proposed in this paper is applicable to other frequency bands of photons, we conducted experiments in the ultraviolet range.

Figure 5(b) summarizes the experimental measurements conducted in the ultraviolet frequency band. The excitation light source is a 15 W ozone-free ultraviolet lamp with stable thermal output performance. The internal cavity temperature variation remains minimal: after 3 minutes of illumination, the cavity temperature rises by only 0.2 °C; after 10 minutes, it increases by 1.3 °C; and after 20 minutes, the total temperature rise reaches 2.2 °C.

Three experimental curves corresponding to different illumination durations exhibit identical overall variation trends but differ in signal amplitude magnitude. Longer pre-shutdown illumination durations lead to a greater accumulation of free dark photons inside the mirror-lined cavity, resulting in a larger resistance drop amplitude in the photoresistor sensing element upon aperture opening.

The ultraviolet light source emits primarily ultraviolet radiation, accompanied by faint pale‑blue‑violet visible light. This mixed spectrum prevents definitive confirmation that all measured signals originate solely from the conversion of ultraviolet photons into dark photons. However, compared with visible light, high-frequency ultraviolet light more readily excites electrons on metal surfaces, thereby triggering the photoelectric effect. This suggests that the detected dark photon signal may predominantly originate from ultraviolet photons.

\par

\section{ Conclusions and Discussion }

\textbf{Conclusions}

\vspace*{3pt}

\textbf{(1)} The independent dark photon theory proposed in this paper introduces a specific microscopic binary structure and establishes a self-consistent theoretical framework distinct from the Standard Model. Define a complete moving photon as a binary composite entity consisting of dark photon and light wave components. Among them, the dark photon embodies the particle nature of light, while the light wave embodies the wave nature of light. This framework reinterprets the fundamental microscopic nature of light and provides logically coherent explanations for a range of optical and astrophysical phenomena, offering a fundamentally new perspective for related fundamental research.

\textbf{(2)} The original dark photon detection methodology and the miniature mirror-lined apparatus developed in this work are fully derived from the core transformation rules of the composite photon model. Multi-band comparative experiments consistently capture anomalous electrical responses that match the theoretical predictions of free dark photon outflow. These observed signals are obtained from our own experimental apparatus and await further independent cross validation by external laboratories.  Compared with large-scale accelerator facilities and high-precision astronomical instruments, the proposed scheme features innovative physical logic, a compact structure, and extremely low cost. It provides a convenient and readily implementable new pathway for exploratory investigations into dark photons.

\textbf{(3)} Although indicative signals consistent with dark photon predictions are repeatedly observed across all experimental groups, this work adopts a rigorous and conservative attitude, concluding only that free dark photons may exist. This prudent statement is justified by the complexity of experimental influencing factors. Full confirmation requires further independent replication by authoritative external laboratories, which constitutes the core objective of the author’s future studies.

\vspace*{3pt}

\noindent\textbf{Discussion}

\vspace*{3pt}

The inferences presented below are purely theoretical deductions derived from the dark photon model constructed in this paper.

\textbf{(1)} The dark photon theory proposed in this paper provides an intuitive and reasonable explanation for the long-debated observer effect. Any observation instrument emits detection signals that interact with complete photons passing through the double slits. This interaction strips bound light-wave components away from complete photon cores, converting complete photons into wave-free dark photons. In the absence of observation equipment, complete photons retain their bound light-wave components and produce characteristic interference fringes on the detection screen via wave superposition. When detection instruments are introduced to track photon trajectories, photon-component separation occurs, suppressing wave interference. In such an observation experiment, when a complete photon loses its wave-like character, the discrete light spots observed on the screen may originate from free dark photons.

\textbf{(2)} Standard cosmological models attribute all observed cosmic redshifts phenomena to the expansion of spacetime. The composite dark photon theory presented in this paper offers another interpretation of astronomical spectral redshift phenomena. Over billions of years of propagation across interstellar space, Photons interact with various substances, such as cosmic dust and nebulae. Short-wavelength photons exhibit a higher probability of having their wave-like properties absorbed during these interactions, converting some photons into free dark photons, thereby losing short-wavelength spectral information. The greater the propagation distance from a celestial object to Earth, the more pronounced the depletion of short-wavelength signal intensity, resulting in an overall macroscopic redshift feature in the received spectrum. This interpretation follows directly from the dark photon theory proposed herein. It describes a global statistical outcome for the universe as a whole, rather than a local effect, and is independent of the specific propagation trajectory of photons.

\textbf{(3)} Independent research groups have reported experimental observations in which photon emission signals appear to exit a medium before the incident photons fully enter it, termed preliminary experimental evidence for negative-time evolution within conventional quantum frameworks [16]. This is regarded as a major scientific discovery in history and holds significant importance as the first experimental evidence supporting negative-time evolution within conventional quantum frameworks. It also provides a new perspective on the dark photon theory proposed in this paper. 

First, at the instant a complete photon interacts with atoms in the medium, its light-wave component separates from its dark photon core within an extremely short time interval. The free dark photon core traverses the material immediately and excites electrons in atoms before the separated light-wave component completes its propagation delay. This time delay results in the observed phenomenon that photons appear to leave the medium before they have fully entered it.

Second, real light sources cannot produce strictly isolated, single-photon emissions; instead, photons are emitted in clusters, with minute differences in arrival times among individual photons. The earliest-arriving dark photon cores excite atoms in the medium and generate measurable emission signals, while a subset of slower photons within the same cluster has not yet crossed the material boundary—resulting in the observed negative-time phenomenon.

\textbf{(4)} Beyond the three representative optical anomalies discussed above, this  dark photon theory also provides coherent, qualitative explanations for other universal optical phenomena, including light scattering effects and the reddish hue of sunset glow, thereby establishing a unified microscopic framework for photon-mediated optical behaviors.

\end{document}